\documentstyle[epsf,rotate]{l-aa}

\newcommand{\kms}{\mbox{\,km\,s$^{-1}$}}

\begin{document}

\title{The rotation speed of the companion star in Aquila X--1 }

\author{T.~Shahbaz$^{1}$ \and J.~Casares$^{2}$ \and P.A.~Charles$^{1}$}

\offprints{T.~Shahbaz (tsh@astro.ox.ac.uk)}

\institute{Department of Astrophysics, Oxford University, Keble Road,
Oxford, OX1 3RH, UK 
\and Instituto de Astrof\'\i{}sica de Canarias
38200 La Laguna. Tenerife, Spain}

\thesaurus{2(08.02.3; 08.06.3; 08.09.2: Aql X--1; 08.14.1)}

\date{Received ?? ?? 1997 / Accepted ?? ?? 1997}

\maketitle

\markboth{T.~Shahbaz et al: Aql X--1}{}


\begin{abstract}
We have obtained medium resolution spectra of the neutron star X-ray
transient Aql X--1 during quiescence. We determine the spectral type of
the companion star to be K1 and also estimate its rotation speed to be
$62^{+30}_{-20}$\kms. By measuring the width of the H$\alpha$ emission
line profile of the accretion disc, we estimate the binary inclination to
$\sim 50^{\circ}$ and also estimate the semi-amplitude of the companion
star's radial velocity curve to be $\sim$170\kms.

\end{abstract}


\section{Introduction}

Aquila X--1(=V1333 Aquilae) is a soft X-ray transient source that shows
X-ray bursts (Koyama et al. 1981; Czerny, Czerny \& Grindlay 1987),
thereby indicating that the compact object is a neutron star. From
quiescent observationss the companion star has been identified to be a
V=19.2 K-type star. The optical counterpart brightens by $\sim$2--5
magnitudes during X-ray outbursts which is interpreted as reprocessing of
radiation in the accretion disk (Thorstensen, Charles \& Bowyer 1978,
hereafter TCB; Canizares, McClintock \& Grindlay 1979; Charles et al.
1980; van Paradijs et al. 1980).

Attempts to find the orbital period have revealed many variations but no
firm period. Watson (1976) reported an unconfirmed 1.3 day X-ray
periodicity during the 1975 outburst. Chevalier \& Ilovaisky (1991) have
obtained an 18.97 hr periodicity from optical photometry during its
active state, which they interpret as being the orbital period.

TCB observed the optical counterpart in its low state. They found the
only significant feature in the spectrum was the Mg$\it b$ 5175\AA\ blend
in absorption; no Balmer lines were evident. They concluded that the
spectrum is characteristic of a K0-3 V star. Shahbaz et al. (1996) also
obtained low resolution spectra of Aql X--1 in quiescence. They too found
the Mg$\it b$ absorption blend but also H$\alpha$ and the Paschen series
in emission. They found by fitting the continuum spectrum that it could
be best described by a K5 star.

Aql X--1 is known to undergo regular X-ray and optical ourbursts on a
timescale of $\sim$1 year (Kaluzienski et al. 1977; Priedhorsky \&
Terrell 1984; Charles et al. 1980) much more frequently than the other
neutron star transient Cen X--4 (McClintock \& Remillard 1990). Recently
the RXTE All Sky Monitor showed Aql X--1 to be have undergone an X-ray
outburst between late January and early March 1997 (Levine \& Thomas
1997). Ilovaisky \& Chevalier (1997) reported that Aql X--1 was optically
in quiescence by 30 March 1997. In this letter we report on our medium
resolution spectra of Aql X--1 obtained in May 1997, when the source was
in quiescence. We were able to determine the spectral type of the
companion star and also its rotation speed.

\section{Observations and data reduction}

We obtained intermediate resolution spectra of Aql X--1 on 13 May 1997
using the 3.9-m Anglo-Australian Telescope at Siding Spring, Australia. A
Tek 1024$^{2}$ CCD attached to the RGO spectrograph was used during all
the observations. The 600V grating was centered at 5190\AA\ giving a
dispersion of 1.55 \AA~pixel$^{-1}$ and a spectral resolution of 3.6\AA\
(FWHM=165 km s$^{-1}$ at H$\alpha$). We took 10 spectra of Aql X--1 each
with an exposure time of 1800 secs starting at 14:58 UT with
interspaced Cu-Ar arc spectra for wavelength calibration. Template field
stars of a variety of spectral types were also observed.

The data reduction and analysis was performed using the Starlink {\sc
figaro} package, the {\sc pamela} routines of K.\,Horne and the {\sc
molly} package of T.\,R.\ Marsh. Removal of the individual bias signal
was achieved through subtraction of the mean overscan level on each
frame. This was acceptable since an examination of the bias frames showed
no significant structure. Small scale pixel-to-pixel sensitivity
variations were removed by multiplying by a flat-field frame prepared
from observations of a tungsten lamp. One-dimensional spectra were
extracted using the optimal algorithm of Horne (1986), and calibration of
the wavelength scale was achieved using 4th order polynomial fits giving
an rms scatter of 1/12th of a pixel (0.13\AA).

\begin{table}
\centering
\small{
\caption{Optimal Subtraction of the Companion Star}
\begin{tabular}{lcccc}\hline
\hline\noalign{\smallskip}
Star 	& Sp. Type & $\chi^{2}$ (DOF=907) & v$\sin\,i$ & $f$ \\ \hline
HR 805  & G8$\sc iii$ 	&   794.2     &  64 	 &  1.05$\pm$0.03 \\
HR 794  & K0$\sc iii$ 	&   779.5     &  62 	 &  1.09$\pm$0.04 \\
HR 807  & K1$\sc iii$ 	&   776.6     &  62 	 &  0.94$\pm$0.03 \\
HR 822  & K2$\sc iii$ 	&   927.2     &  66 	 &  0.77$\pm$0.03 \\
HR 872  & K3$\sc iii$ 	&   900.6     &  66 	 &  0.60$\pm$0.02 \\
HR 871  & K4$\sc iii$ 	&  1040.4     &  36 	 &  0.55$\pm$0.02 \\
HR 851  & K5$\sc iii$ 	&   861.6     &  50 	 &  0.70$\pm$0.02 \\ \hline
\noalign{\smallskip}
\hline
\end{tabular}
}
\end{table}

\begin{figure*}
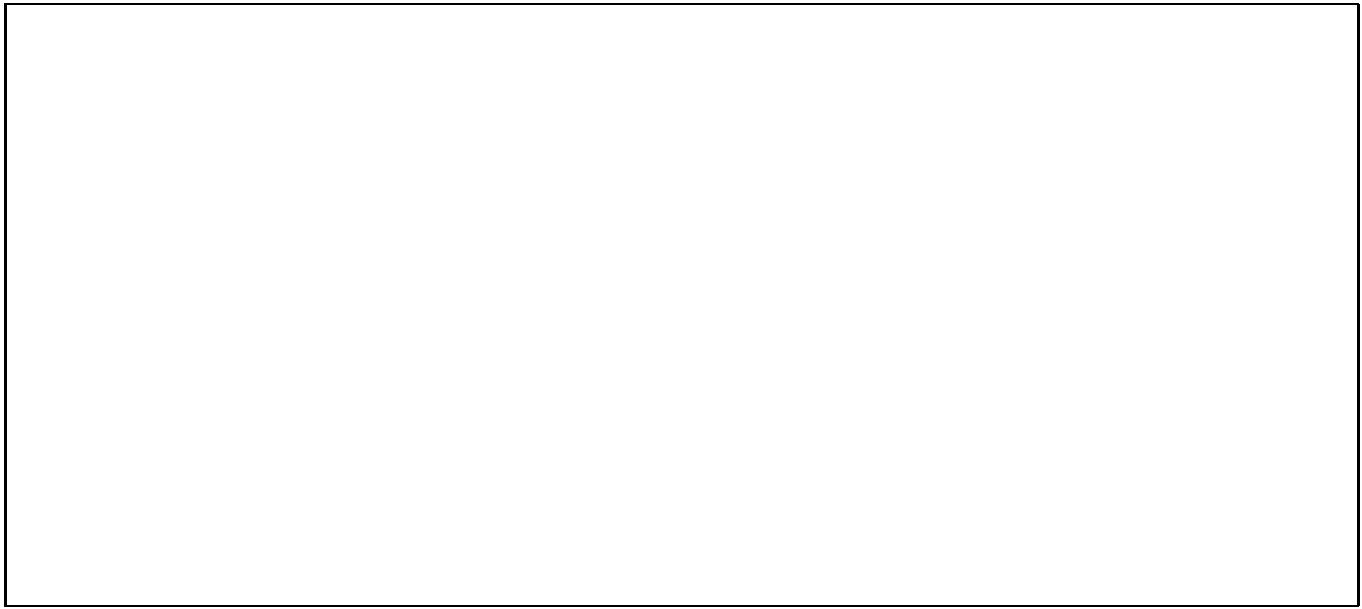

  \picplace{8cm}
  \caption{ The results of the optimal subtraction. From top to bottom:
  the variance-weighted average spectrum of Aql X--1, the template K1$\sc
  iii$ star (HR 907) broadened by 62\kms, the residual spectrum of Aql
  X--1 after subtracting the template star times $f$=0.94. The spectra
  have been normalized and shifted vertically for clarity. The right
  inset shows a close-up of the smoothed double-peaked H$\alpha$ emission
  line profile in the Aql-X--1 spectrum. The left inset shows a close-up
  of the H$\alpha$ residual spectrum with FWHM=816$\pm$40\kms.}
\end{figure*}

\section{The spectral type and rotational broadening of the companion star}

We determine the spectral type of the companion star by minimizing the
residuals after subtracting different template star spectra from the
Doppler-corrected average spectrum. This method is sensitive to the
rotational broadening $v\sin\,i$ and the fractional contribution of the
companion star to the total flux ($f$; 1-$f$ is the ``veiling factor'').

First we determined the velocity shift of the individual spectra of Aql
X--1 with respect to each template star spectrum by the method of
cross-correlation (Tonry \& Davis 1979). The Aql X--1 spectra were then
interpolated onto a logarithmic wavelength scale (pixel size 80 \kms)
using a $\sin\,x/x$ interpolation scheme to minimize data smoothing
(Stover et al. 1980). No significant velocity variations were observed
(see section 4), so we averaged the Aql X--1 spectra to the rest frame of
the template star.

In order to determine the rotational broadening $v\sin\,i$ we follow the
procedure described by Marsh, Robinson \& Wood (1994). Basically we
subtracted a constant representing the fraction of light from the
template star, multiplied by a rotationally broadened version of that
template star. We broadened the template star spectrum from 0 to 200
km~s$^{-1}$ in steps of 1 km~s$^{-1}$ using the Gray rotation profile
(Gray 1976). We then performed an optimal subtraction between the
broadened template and averaged Aql X--1 spectra. The optimal subtraction
routine adjusts the constant to minimize the residual scatter between the
spectra. The scatter is measured by carrying out the subtraction and then
computing the $\chi^{2}$ between this and a smoothed version of itself.
The constant, $f$, represents the fraction of light arising from the
template spectrum, i.e. the secondary star. The optimal values of
$v\sin\,i$ and $f$ are obtained by minimising $\chi^{2}$ (see Table 1).
The above analysis was performed in the spectral range 5120--6710 \AA\,
excluding the H$\alpha$ emission line and the interstellar feature at
5190 \AA. A linear limb-darkening coefficient of 0.76 was used (Al-Naimiy
1978) appropriate for 6000 \AA\ and an effective temperature of 4500 K
(typical for a K star). We also performed the above analysis using zero
and full limb-darkening and only obtained a change in $v\sin\,i$ of at
most 4 \kms.

From Table 1 it can be seen that the minimum $\chi^{2}$ occurs at
spectral type K1 with a $v\sin\,i$ of 62$^{+30}_{-20}$ \kms (1-$\sigma$)
and the companion star contributing about 94\% to the observed flux at
$\sim$ 6000 \AA. Fig. 1 shows the results of the optimal subtraction and
the $\chi^{2}$ -- $v\sin\,i$ plot is shown in Fig. 2. All the template
stars used in the above analysis were of luminosity class $\sc iii$. The
effects of using different luminosity class spectra in the analysis was
estimated by comparing the absorption line flux density of stars of
different spectral type and class. We find using template stars of
spectral type K1, K4 and K7 and luminosity class $\sc v$ and $\sc iii$,
the flux density of the stars of different luminosity class (but the same
spectral type) are only different at the 1 percent level. Therefore, the
value we obtain for $f$ in the above analysis is not dependant on the
luminosity class of the template star used.

In order to verify the validity of the minimum in the $\chi^{2}$ --
$v\sin\,i$ plots we performed the same analysis on a template star. The
broadening in the template star spectrum is dominated by the instrumental
broadening. To this template star we added noise to produce a spectrum of
comparable quality to our Aql X--1 data. The above broadening and optimal
subtraction procedure was then repeated, The results are shown in Fig. 2,
where the minimum $\chi^{2}$ occurs when no broadening is used. This is
exactly as expected since the template star spectrum must be dominated by
instrumental broadening. Therefore we are confident that the minimum
$\chi^{2}$ for the Aql X--1 spectrum is real, although the errors in the
$v\sin\,i$ determination are large.

\section{The Aql X--1 spectrum}

In Fig. 1 we show the variance-weighted average of the Aql X--1 spectra,
which has a signal-to-noise ratio of about 25 in the continuum. The most
noticeable feature is double peaked H$\alpha$ emission which has a broad
base (FWZI$\sim$2000\kms). We tried to see if there was any velocity
variation in the absorption lines by cross-correlating the individual Aql
X--1 spectra with a template star (HR 807), but found no significant
variation over the 5 hr baseline. Therefore we quote the mean
heliocentric systemic velocity obtained from the averaged Aql X--1
spectrum of 57$\pm$21\kms (the template star used, HR 807, had a
heliocentric velocity of 15$\pm$16\kms). Again from the averaged Aql X--1
spectrum the equivalent width of H$\alpha$ is 5.3$\pm$0.3\AA\, with a
mean heliocentric velocity of 99$\pm$30\kms, obtained using a single
Gaussian fit.

\section{Discussion}

\subsection{The binary inclination}

After subtracting the spectrum of the companion star from the Aql X--1
spectrum, the most noticable feature is the single peaked H$\alpha$
emission line (presumably arising from the accretion disc) with a FWHM
velocity of 816$\pm$40\kms (see Fig. 1). This is in contrast to the
observed Aql X--1 spectrum which shows a double-peaked profile, implying
that the core of the double-peaked profile is due to the H$\alpha$
absorption line arising from the secondary star.

We can compare the accretion disc H$\alpha$ emission line in Aql X--1
with that in Cen X--4 since both have neutron star primaries and probably
similar orbital periods. The width of the disc H$\alpha$ emission line in
Cen X--4 is 687$\pm$20\kms (FWHM; Casares et al. 1997) whereas it is
broader in Aql X--1, which suggests that the binary inclination of Aql
X--1 may be higher than that of Cen X--4. If we assume Keplerian motion
in the accretion disc and that the mass of the compact object, and
orbital period of the two systems are comparable, then the velocities in
the disc simply scale with $\sin\,i$ (Frank, King \& Raine 1992).
Using the velocity widths given
above for Aql X--1 and Cen X--4 and the binary inclination for Cen X--4
(40$^{\circ}$; Shahbaz, Naylor \& Charles 1993), we estimate $i\sim
50^{\circ}$ for Aql X--1.

It should however, be noted that the assumption of Keplerian velocities
at the outer edge of the disc may not hold (Orosz et al. 1994). For a
system at such an inclination one might have expected to see a
double-peaked emission line profile, arising from an accretion disk
viewed at high inclination. However, there are probably other sources of
H$\alpha$ emission from the system which contaminate the disc profile,
such as from the heated face of the secondary star or from the bright
spot. Also note the poor resolution and signal-to-noise of our data,
therefore we cannot rule out a high binary inclination on the lack of a
double-peaked profile alone.

\begin{figure}
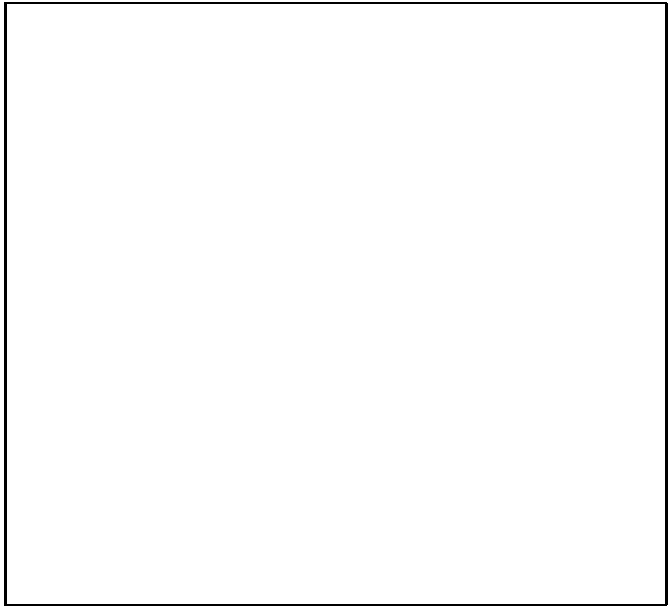

  \picplace{8cm}
  \caption{The $\chi^{2}$ -- $v\sin\,i$ analysis plot for Aql X--1 (shown
  as the histogram) using the template star HR 907. The minimum
  $\chi^{2}$ ($\chi^{2}_{min}$) is found with a rotational broadening of
  62 \kms. The 68\% confidence level ($\chi^{2}_{min}$+1) is also marked
  as the dashed horizontal line. Also shown is the result for simulated
  data using a template star shown as the solid line. Note that
  $\chi^{2}_{min}$ occurs at zero broadening, which implies that the
  minimum in the Aql X--1 $\chi^{2}$ -- $v\sin\,i$ plot is real. The
  right abscissa refers to the template star $\chi^{2}$ values.}
\end{figure}

\subsection{The $K_{2}$ velocity}

If the orbital period is 18.97 hr and the secondary star fills its Roche
lobe, then the density of the secondary star is $\rho$=0.3 g~cm$^{-3}$. A
K1 main sequence star would have $\rho$=1.9 g~cm$^{-3}$., implying that
the radius of the secondary star must be about twice that of a main
sequence star in order for it to fill its Roche lobe. It is therefore
likely to be evolved, similar to the secondary star in Cen X--4 (Shahbaz,
Naylor \& Charles 1993). If we believe the estimate to the binary
inclination to be $\sim 50^{\circ}$, and that both systems have similar
mass compact objects, then one can estimate $K_{2}$ for Aql X-1 ($K_{2}$
scales with $\sin\,i$ c.f. the mass function equation). Using $K_{2}$=146
\kms (McClintock \& Remillard 1990) and $i=40^{\circ}$ for Cen X--4 and
$i=50^{\circ}$ for Aql X--1 we estimate $K_{2}$ for Aql X--1 to be
$\sim$170\kms.

\section{Conclusions}

Using medium resolution optical spectra, we have determined the spectral
type of the companion star in the neutron star X-ray transient system Aql
X--1, to be a K1 star. By optimally subtracting different broadened
versions of the companion star spectrum from the average Aql X--1
spectrum we determine the rotational broadening of the companion star to
be $62^{+30}_{-20}$\kms, with a contamination of $\sim$ 6\% to the
observed flux at 6000\AA. We also estimate the binary inclination to be
$\sim 50^{\circ}$ by measuring the width of the accretion disc H$\alpha$
emission line, and $K_{2}$ to be $\sim$ 170 \kms.

\section*{Acknowledgments}

We are grateful to all the staff at the AAT, Siding Spring in particular
our telescope operator Jonathan Pogson. The data reduction was carried
out on the Oxford Starlink node, and the figures were plotted using the
$\sc ark$ software.

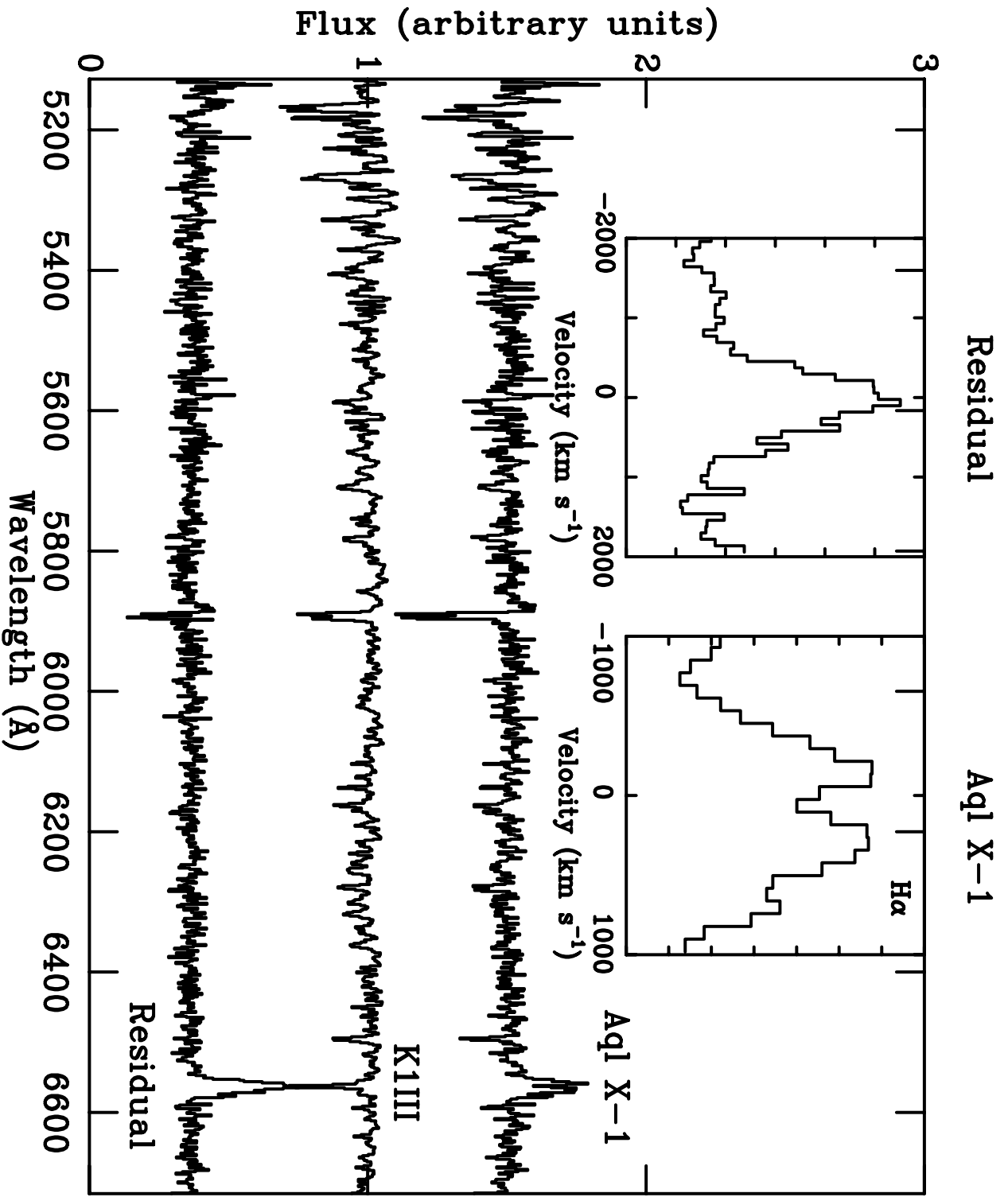
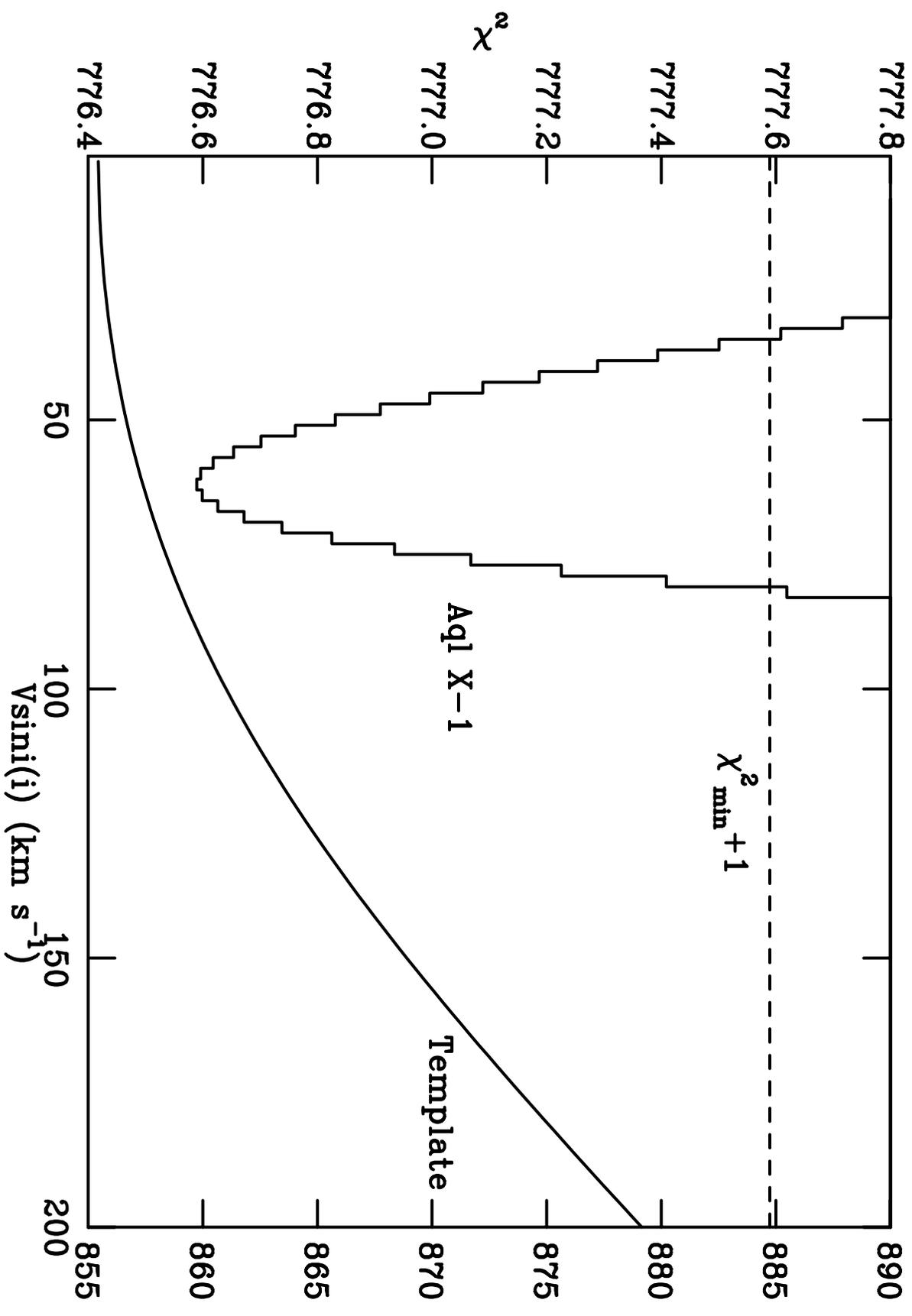
\end{document}